\documentclass[a4paper,11pt]{article}
\usepackage{amsmath,bm,amssymb,color,mathtools}

\usepackage{booktabs}
\usepackage{multirow}

\usepackage{graphicx}
\usepackage{jheppub} % for details on the use of the package, please
\usepackage{xspace}

\renewcommand{\tabcolsep}{2em}

\DeclareRobustCommand{\ensuremathrm}[1]{\ensuremath{\mathrm{#1}}}
\DeclareRobustCommand{\rd}{\ensuremathrm{d}}  % differential operator
\DeclareRobustCommand{\rT}{\ensuremathrm{T}}  % transverse
\DeclareRobustCommand{\rL}{\ensuremathrm{L}}  % longitudinal
\DeclareRobustCommand{\rR}{\ensuremathrm{R}}  % ren
\DeclareRobustCommand{\rF}{\ensuremathrm{F}}  % fact
\DeclareRobustCommand{\PZ}{{\ensuremathrm{Z}}\xspace}
\DeclareRobustCommand{\Pl}{{\ensuremath{\ell}}\xspace}

\DeclareRobustCommand{\Plp}{{\ensuremath{\ell^+}}\xspace}
\DeclareRobustCommand{\Plm}{{\ensuremath{\ell^-}}\xspace}
\DeclareRobustCommand{\Plpm}{{\ensuremath{\ell^\pm}}\xspace}

\DeclareRobustCommand{\GeV}{\ensuremathrm{GeV}\xspace}
\DeclareRobustCommand{\TeV}{\ensuremathrm{TeV}\xspace}

\DeclareRobustCommand{\jet}{\text{jet}\xspace}
\DeclareRobustCommand{\cut}{\text{cut}\xspace}
\DeclareRobustCommand{\acop}{\text{acop}\xspace}
\DeclareRobustCommand{\depart}{\text{depart}\xspace}
\DeclareRobustCommand{\muR}{\ensuremath{\mu_{\rR}}\xspace}

\DeclareRobustCommand{\muF}{\ensuremath{\mu_{\rF}}\xspace}

\DeclareRobustCommand{\order}[1]{\ensuremath{{\mathcal{O}\!\left(#1\right)}}\xspace}
\DeclareRobustCommand{\alphas}{\ensuremath{\alpha_{\mathrm{s}}}\xspace}

\DeclareRobustCommand{\thstar}{\ensuremath{\theta^{*}_{\eta}}\xspace}
\DeclareRobustCommand{\phistar}{\ensuremath{\phi^{*}_{\eta}}\xspace}
\DeclareRobustCommand{\phistardep}{\ensuremath{\phi^{*}_{\eta,\depart}}\xspace}
\DeclareRobustCommand{\phistarcut}{\ensuremath{\phi^{*}_{\eta,\cut}}\xspace}
\DeclareRobustCommand{\phistarlarge}{0.051}
\DeclareRobustCommand{\ptz}{\ensuremath{p^{\PZ}_{\rT}}\xspace}
\DeclareRobustCommand{\ptzdep}{\ensuremath{p^{\PZ}_{\rT,\depart}}\xspace}
\DeclareRobustCommand{\ptzcut}{\ensuremath{p^{\PZ}_{\rT,\cut}}\xspace}
\DeclareRobustCommand{\yz}{\ensuremath{y^{\PZ}}\xspace}
\DeclareRobustCommand{\mll}{\ensuremath{m_{\Pl\Pl}}\xspace}
\DeclareRobustCommand{\NNLOJET}{{\normalfont\textsc{NNLOjet}}\xspace}
\DeclareRobustCommand{\vecpt}[1]{\ensuremath{\vec{p}^{\,#1}_{\rT}}\xspace}
\DeclareRobustCommand{\gapprox}{\gtrsim}
\allowdisplaybreaks

\begin{document}

\preprint{IPPP/16/74, ZU-TH 36/16}

\title{NNLO QCD corrections for Drell--Yan $\ptz$ and $\phistar$ observables at the LHC}

\author{A.\ Gehrmann--De Ridder$^{a,b}$, T.\ Gehrmann$^{b}$, E.W.N.\ Glover$^c$, A.\ Huss$^{a}$, T.A.\ Morgan$^c$}

\affiliation{
$^a$Institute for Theoretical Physics, ETH, CH-8093 Z\"urich, Switzerland\\
$^b$Department of Physics, University of Z\"urich, CH-8057 Z\"urich, Switzerland\\
$^c$Institute for Particle Physics Phenomenology, Department of Physics, University of Durham, Durham, DH1 3LE, UK}
\emailAdd{gehra@phys.ethz.ch}
\emailAdd{thomas.gehrmann@uzh.ch}
\emailAdd{e.w.n.glover@durham.ac.uk}
\emailAdd{ahuss@phys.ethz.ch}
\emailAdd{t.a.morgan@durham.ac.uk}

\abstract{
Drell--Yan lepton pairs with finite transverse momentum are 
produced when the vector boson recoils against (multiple) parton emission(s), and is determined by QCD dynamics.  At small transverse momentum, the fixed order predictions break down due to the emergence of large logarithmic contributions.  This region can be studied via the $\ptz$ distribution constructed from the energies of the leptons, or through the $\phistar$ distribution that relies on the directions of the leptons.  For sufficiently small transverse momentum, the $\phistar$ observable can be measured experimentally with better resolution. 
We study the small $\ptz$ and $\phistar$ distributions up to next-to-next-to-leading order (NNLO) in perturbative QCD. 
We compute the $\phistar$ distributions for the fully inclusive production of lepton pairs via $\PZ/\gamma^*$ to NNLO and normalise them to the NNLO cross sections for inclusive $\PZ/\gamma^*$ production. 
We compare our predictions with the $\phistar$ distribution measured by the ATLAS collaboration during LHC operation at 8 TeV. We find that at moderate to large values of $\phistar$, the NNLO effects are positive and lead to a substantial improvement in the theory--data comparison compared to next-to-leading order (NLO). At small values of $\ptz$ and $\phistar$, the known large logarithmic enhancements emerge through and we identify the region where resummation is needed.  We find an approximate relationship between the values of $\ptz$ and $\phistar$ where the large logarithms emerge and find perturbative consistency between the two observables.}

\maketitle

%------------------------------------------------
\section{Introduction}
\label{sec:intro}
%------------------------------------------------

%-Motivation DY
The production of $\PZ$-bosons which subsequently decay into a pair of leptons is a Standard Model benchmark 
process at hadron colliders. It occurs with a large rate and, due to its clean final state signature, can be measured very accurately with small experimental uncertainties. It has been studied extensively at the LHC by the ATLAS~\cite{ptzATLAS7TeV,ptzATLAS}, CMS~\cite{ptzCMS7TeV,ptzCMS} and LHCb~\cite{ptzLHCb} experiments. 

When combined with precise theoretical predictions for related observables, there is the potential for accurate determinations of fundamental parameters of the theory. In particular, the transverse momentum distribution of the $\PZ$-boson has been one of the most studied observables. The high sensitivity of the $\ptz$ spectrum to the distribution of gluons in the proton makes it a key observable for constraining parton distribution functions (PDF's).

For inclusive $\PZ$-production, restricting ourselves to the framework of QCD, corrections at next-to-next-to-leading order (NNLO) are available~\cite{dyNNLO,dynnlo,fewz,vrap} and the present state of the art is obtained by combining the NNLO QCD corrections with a resummation of next-to-next-to-leading logarithmic effects (NNLL)~\cite{dyresum}. This combination is necessary to predict the transverse momentum distribution of the $\PZ$-boson at small $\ptz$.
 In this region, large logarithmic corrections of the form $\ln^n(\ptz/\mll)$ appear at each order in the perturbative expansion in $\alphas$, spoiling the convergence of the fixed-order predictions. 

% recoil and ptz paper
The transverse momentum of the $\PZ$-boson is caused by the emission of QCD radiation from the initial state partons. As a consequence, fixed order predictions at $\order{\alphas^2}$ in perturbative QCD, which are NNLO accurate for the inclusive cross section correspond only to NLO accurate predictions for the transverse momentum distributions. 
At high values of $\ptz$, namely above $20~\GeV$, both ATLAS and CMS observed a tension between the NLO predictions and their measurements of the $\ptz$ distributions presented in the form of fiducial cross sections for a restricted kinematical range of the final state leptons. 
Motivated by this observation, in a recent paper~\cite{PTZus}, we have used the parton-level event generator \NNLOJET, as described in~\cite{ZJNNLOus},  to predict the $\PZ$-boson distributions 
at large transverse momentum to NNLO  accuracy. 
We  computed the fiducial cross section for the production of a $\PZ$-boson at finite transverse momentum fully inclusively on the hadronic final state. We found that when the $\ptz$ distribution is normalised to the relevant di-lepton cross section, the NNLO predictions yield an excellent agreement with the measured distributions at $\sqrt{s}= 8~\TeV$ from both ATLAS and CMS over a large range of $\ptz$ values above the selected cut of $\ptz= 20~\GeV$. Given the importance of a precise determination of the $\ptz $ spectrum for phenomenology, it is also crucial to have a thorough probe of the low $\ptz$ domain as well. We therefore use the \NNLOJET code to make predictions in the low transverse momentum region.  As expected, the fixed order description will fail at sufficiently small $\ptz$, but it is also interesting to see exactly where this happens. 
% In particular, we will show that the NNLO fixed order perturbative description extends to significantly lower $\ptz$ than at NLO.

% motivation for PTZ sensitive observables and phistar
In the small $\ptz$ region, the precision of direct measurements of the $\ptz$ spectrum using the standard $\ptz$ variable is limited by the experimental resolution on $\ptz$ itself, and in particular on the resolution of the magnitude of the transverse momenta of the individual leptons entering $\ptz$.
To probe the low $\ptz$ domain of $\PZ/\gamma^*$ production an alternative angular variable,
$\phistar$, has been proposed~\cite{banfidefphistar} which minimises the impact of these experimental uncertainties. It is defined by
\begin{equation} 
  \label{eq:phistardef}
  \phistar\equiv \tan\left(\frac{\phi_{\acop}}{2}\right) \cdot \sin (\thstar) .
\end{equation} 
In this definition, the acoplanarity angle is 
\begin{equation}
  \label{eq:acop}
  \phi_{\acop}\equiv \pi -\Delta\phi 
  \equiv 2 \arctan \left(\sqrt {\frac{1 + \cos \Delta \phi}{ 1 -\cos \Delta \phi}}\right) ,
\end{equation}
where $\Delta\phi$ is the azimuthal angle between the two leptons. 
The angle $\thstar$ is the scattering angle of the leptons with respect to the proton beam direction in the reference frame that is boosted along the beam direction such that the two leptons are back-to-back in the $(r,\theta)$ plane.
It is explicitly given by
\begin{equation}
  \cos(\thstar)\equiv \tanh\left(\frac{\eta^{\Plm} -\eta^{\Plp}}{2}\right) ,
\end{equation}
where $\eta^{\Plm}$ and $\eta^{\Plp}$ are the pseudorapidities of the negatively and positively charged leptons respectively. 

The variable $\phistar$ measures the ``deviation from back-to-backness'' (acoplanarity) in the transverse plane and therefore 
vanishes at Born level where the azimuthal angle between the two leptons $\Delta\phi$ is exactly equal to $\pi$. 
Non-zero values of $\phistar$ are produced by the same mechanism that generates non-zero $\ptz$, namely a recoil against hadronic emission from the partonic initial states.  As a consequence, the $\phistar$ distribution probes the same type of physics as the transverse momentum distribution. 
As we shall see in section~\ref{sec:phistardef},  in the small $\ptz$ limit, $\phistar$ is explicitly related to $\ptz/\mll$ where $\mll$ is the invariant mass of the lepton pair.
Furthermore, $\phistar$ is positive by construction and depends exclusively on the directions (rather than the magnitudes) of the lepton momenta. As the directions of the leptons are considerably better measured than their transverse momenta, analysing the low $\ptz$ region with this angular kinematical variable $\phistar$ has the potential to increase the accuracy of the measurements and opens up the possibility of making more stringent tests of the theoretical predictions for both observables. 

% data and comparison data -theory  
So far, the $\phistar$ distribution and related observables have been studied at the Tevatron by the D0 collaboration~\cite{D0phistar} and by the ATLAS Collaboration at the LHC at $\sqrt{s}=7~\TeV$~\cite{atlasphistar7Tev} and $\sqrt{s}= 8~\TeV$~\cite{ptzATLAS} and very recently by the LHCb Collaboration at $13~\TeV$~\cite{lhcb13}. The $7~\TeV$ data from ATLAS have been compared with theoretical predictions from the Monte Carlo program RESBOS~\cite{Resbos} which includes NLO fixed order corrections, resummation, and non-perturbative effects. The $7~\TeV$ measurements have also been compared to the fixed order NNLO inclusive $\PZ$ prediction of FEWZ~\cite{fewz} and to a matched NLO+NNLL resummed computation \cite{BanfiphistarLHC}, and to a NNLL+NNLO computation including leptonic decay~\cite{Catani:2015vma}.

Even above $\phistar\sim 0.1$, a relatively large value of $\phistar$, the NNLO predictions obtained with FEWZ undershoot the data by about ten percent.
This is not a surprise given that although these predictions are NNLO accurate for the inclusive cross section, they are only NLO accurate for the $\phistar$ distribution as for the $\ptz$-spectrum. 
The theoretical predictions for the NNLL resummed calculation matched to NLO fixed order 
show reasonable agreement with the data, but with large theoretical uncertainties.
The $\phistar$ measurements of ATLAS at $8~\TeV$~\cite{ ptzATLAS} have thus so far only been compared 
with results obtained from parton shower Monte Carlo programs~\cite{MC} and RESBOS.

% Aim of the present paper%
It is the purpose of this paper to explore the production of lepton pairs at low (but non-zero) transverse momentum at NNLO for both the $\ptz$ and $\phistar$ distributions.  
In particular, we will make the first NNLO accurate predictions for observables related to  
$\phistar$ at non-vanishing $\phistar$ and make a direct comparison with 
the $8~\TeV$ ATLAS $\ptz$ and $\phistar$ data \cite{ptzATLAS} by using the same fiducial cuts for the leptons as those used in the experimental measurements. 
The NNLO predictions are obtained using our parton-level code \NNLOJET, designed to compute NNLO corrections to observables related to $\PZ+\jet$ production~\cite{ZJNNLOus}, by being completely inclusive on the QCD radiation (i.e.\ dropping the requirement of observing a jet) and applying a low cut on either $\ptz$ or $\phistar$. We make predictions for all but the first bin in $\ptz$ ($0$--$2~\GeV$) or $\phistar$ ($0$--$0.004$) where the fixed order prediction for the distribution formally diverges.

As expected, in the very low transverse momentum domain, reliable theoretical predictions can only be provided through the resummation of large logarithms of the form $\ln^n(\ptz)$ or $\ln^n(\phistar)$ to all orders in perturbation theory. We shall see these large logarithmic effects emerge from the fixed order distributions shown in section~\ref{sec:results}. 
Nevertheless, we find that the inclusion of the NNLO corrections to the normalised distributions 
accurately describe the data over a wide range of values of $\ptz$ and $\phistar$.
% In particular, we will show that the NNLO fixed order predictions accurately describe the data to much lower values of $\ptz$ and $\phistar$ than the corresponding predictions obtained at NLO. 
% By studying the form of the large logarithms we find an approximate relationship between the values of $\ptz$ and $\phistar$ where the fixed order distributions start to break down and find consistency between the two observables. 
Using an aproximate kinematic relation between $\ptz$ and $\phistar$ valid in the low-$\ptz$ regime as a starting point, we examine the breakdown points of the fixed-order predictions and find consistency between the two observables.

%------------------------------------------------
\section{Kinematics of the angular variable \texorpdfstring{$\phistar$}{phi*} in the low-\texorpdfstring{$\ptz$}{ptz} regime}
\label{sec:phistardef}
%------------------------------------------------

\begin{figure}[t]
  \centering
  \includegraphics[width=6.0cm]{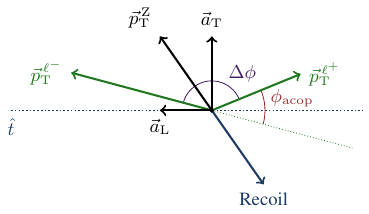}
  \caption{Illustration of the variables needed in the decomposition of the transverse momentum vector of the di-lepton system: $\vecpt{\PZ}$ and the angles entering the definition of $\phi^{*}$ (see text). The hadronic recoil is expected to have an equal and opposite transverse momentum vector to that of the di-lepton system vector $\vecpt{\PZ}$.}
  \label{fig:phistar}
\end{figure}

Let us first consider how the variable $\phistar$ defined in Eq.~\eqref{eq:phistardef} is related to the transverse momentum of the $\PZ$ boson for events where the leptons are separated by an angle $\Delta\phi$ which is greater than $\pi/2$, cf.\ Refs.~\cite{banfidefphistar,wyattdefphistar}. 
This type of event is illustrated in Fig.~\ref{fig:phistar}. 
The transverse momentum vector of the di-lepton system, $\vecpt{\PZ}$, can be decomposed into orthogonal components with respect to an event axis in the plane transverse to the beam direction.
The lepton thrust axis  is defined by a unit vector $\hat{t}$, 
\begin{equation}
  \hat{t} = \frac{\left(\vecpt{\Plm} - \vecpt{\Plp}\right)}{| \vecpt{\Plm} - \vecpt{\Plp} |} ,
\end{equation}
where $\vecpt{\Plm}$ and $\vecpt{\Plp}$ are the lepton momentum vectors in the transverse plane.
The longitudinal and transverse components of $\vecpt{\PZ}$ are denoted by
$\vec{a}_\rL$ and $\vec{a}_\rT$. Their respective magnitudes denoted by $a_\rL$ and $a_\rT$ are related to $\ptz$ by
\begin{equation}
  \label{eq:ptzalat}
  \ptz=\sqrt{a_\rT^2 +a_\rL^2} .
\end{equation}

We are particularly interested in the relation between the variables $\ptz$ and $\phistar$ when the $\PZ$ boson transverse momentum is small.  
In the low $\ptz$ limit we are in ``quasi-Born-like" kinematics where the two leptons are almost back-to-back in the transverse plane and $\Delta\phi \approx \pi$.
The magnitudes of the lepton momenta then satisfy 
\begin{equation}
  \label{eq:ptlapprox}
  p_\rT^{\Plm} \approx p_\rT^{\Plp} \equiv p_\rT^{\Pl} \gg \ptz
\end{equation} 
and the scattering angle behaves as
\begin{equation}
  \sin(\thstar) \approx \frac{2 p_\rT^{\Pl}}{\mll} .
\end{equation}
Furthermore, in this limit $a_\rL$ and $a_\rT$ are approximately given by
\begin{eqnarray}
  \label{eq:aTdef}
  a_\rL^2 &\equiv&  (\ptz)^2 \cos^2  \alpha \approx (p_\rT^{\Plm} - p_\rT^{\Plp})^2 , \\
  a_\rT^2 &\equiv&  (\ptz)^2 \sin^2 \alpha \approx 2 (p_\rT^{\Pl})^2 (1 + \cos \Delta \phi) ,
\end{eqnarray}
where $\alpha$ is the angle between $\vecpt{\PZ}$ and $\hat{t}$.
Using the definition of \phistar in Eq.~\eqref{eq:acop}, in the small $\ptz$ limit (i.e.\ $\Delta \phi \approx \pi$), we arrive at
\begin{equation}
  \label{eq:phistaraT}
  \phistar \approx \frac{a_\rT}{\mll} ,
\end{equation}
where we have used 
\begin{equation}
  \tan \left(\frac {\phi_{\acop}}{2}\right) 
  \approx 
  \frac{a_\rT}{2 p_\rT^{\Pl}} .
\end{equation} 

From Eq.~\eqref{eq:phistaraT}, it is clear that in the small $\ptz$ region, the $\phistar$ distribution probes the same physics as the $a_\rT$ and $\ptz$ distributions (since $a_\rT=\ptz \sin \alpha$).   
% In section~\ref{sec:low-phi} we will further establish a more direct relation between $\phistar$ and $\ptz$ by examining the arguments of the logarithms that appear in the respective resummation formulae.

%------------------------------------------------
\section{Numerical results}
\label{sec:results}
%------------------------------------------------

The results presented in this section are based on the calculation of Ref.~\cite{ZJNNLOus}, where the NNLO QCD corrections to $\PZ+\jet$ production were computed using the antenna subtraction formalism~\cite{ourant} to isolate the infrared singularities in the different $\PZ$-boson-plus-jet contributions. Our calculation is 
implemented in a newly developed parton-level Monte Carlo generator \NNLOJET. This program
provides the necessary infrastructure for the antenna subtraction of hadron collider processes at NNLO and performs the integration of all contributing subprocesses at this order. Components of it have also been used in other NNLO QCD calculations~\cite{eerad3,nnlo2j,nnlohj,nnlott,ZJNNLOus,PTZus,nnlodis} using the antenna subtraction method. Other processes can be added to \NNLOJET provided the matrix elements are 
available. 

In Refs.~\cite{PTZus,PTZproceedings}, we showed that \NNLOJET can be used to predict the $\PZ$-boson $p_\rT$ spectrum to genuine NNLO accuracy by dropping the requirement of observing a jet and instead imposing a finite cut on the transverse momentum $\ptz > \ptzcut$.  These predictions are therefore completely inclusive on the QCD radiation and depend only on the fiducial cuts applied to the leptons. 
This calculation is extended in this work to substantially lower values of $\ptz$. 
Since the $\phistar$ variable is related to $\ptz$, we can equally use \NNLOJET to compute the the first NNLO accurate predictions for $\phistar > \phistarcut$ in the same way.  
By going to low values of $\ptz$ or $\phistar$, one starts to  resolve 
the N$^3$LO infrared singularity at $\ptz=0$ or $\phistar=0$ which
presents a challenge for any NNLO subtraction or slicing method.

%------------------------------------------------
\subsection{Calculational setup} 
\label{sec:setup}
%------------------------------------------------

The ATLAS collaboration measured~\cite{ptzATLAS} the $\ptz$ and $\phistar$ distributions at 8 TeV by applying fiducial acceptance cuts on the leptons:
\begin{align}
	\label{eq:cuts}
	\lvert \eta^{\Plpm} \rvert & < 2.4 , &
	p_{\rT}^{\Plpm} & > 20~\GeV , &	
	46~\GeV < \mll & < 150~\GeV , &
	\lvert \yz \rvert & < 2.4 ,
\end{align}
where $y^{\PZ}$ denotes the rapidity of the lepton pair. We apply the same cuts in our calculation. 
As discussed above, a non-zero cut $\phistar > \phistarcut$ or $\ptz > \ptzcut$ has to be applied in order to render the $\PZ+\jet$ calculation infrared safe. 
We choose the value $\phistarcut = 0.004$ for the $\phistar$ distribution and
$\ptzcut=2$~GeV for the $\ptz$ distribution, each time corresponding to the upper edge of the first bin in the ATLAS data set.

For our numerical computations, we use the NNPDF3.0 parton distribution functions~\cite{nnpdf} with the value of $\alphas(M_{\PZ})=0.118$ at NNLO, and $M_{\PZ}=91.1876~\GeV$. 
Note that we systematically use the same set of PDFs and the same value of $\alphas(M_{\PZ})$ for the NLO and NNLO predictions. 
The factorisation and renormalisation scales are chosen dynamically on an event-by-event basis using the central scale
\begin{equation}
	\label{eq:scale}
	\mu_0 \equiv \sqrt{\mll^2 + (\ptz)^2},
\end{equation}
where $\mll$ is the invariant mass of the final state lepton pair. 
The theoretical uncertainty is estimated using the standard 7-point scale variation, i.e.\ varying $\muR$ and $\muF$ independently about $\mu_0$ by multiplicative factors in the range $[1/2, 2]$ while retaining $1/2 < \muR/\muF < 2$.
We restrict our discussion to normalised distributions which are much more precisely determined due to the fact that systematic errors such as the luminosity uncertainties cancel in the ratio.
To this end, we use the implementation of the NNLO QCD corrections to inclusive $\PZ/\gamma^*$ production available in \NNLOJET to compute the fiducial cross section for the respective bins.

The measurement of the $\ptz$  and $\phistar$  distributions in Ref.~\cite{ptzATLAS} are performed multi-differentially with additional binning in the invariant mass ($\mll$) and the rapidity ($y^\PZ$) of the lepton pair.
The invariant-mass range of Eq.~\eqref{eq:cuts} is divided into three mass bins: 
the \PZ resonance bin containing the $\PZ$-boson peak ($\mll \in [66,116]~\GeV$) and two off-resonance bins covering the low-mass ($\mll \in [46,66]~\GeV$) and high-mass ($\mll \in [116,150]~\GeV$) regions. 
Each mass bin is further subdivided into equal-sized rapidity bins---six for the resonance region and three for each of the off-resonance regions.
The ATLAS measurement of the $\ptz$ distribution~\cite{ptzATLAS} further extends to lower invariant masses, with three more bins in the range below $46~\GeV$. For those bins, results are provided only for $\ptz>45~\GeV$, which have been studied to NNLO accuracy already in Ref.~\cite{PTZus}.

%------------------------------------------------
\subsection{The transverse momentum distribution at low \texorpdfstring{$\ptz$}{ptz}}
\label{sec:low-ptz}
%------------------------------------------------

The ATLAS measurement of the transverse momentum distribution~\cite{ptzATLAS} starts at vanishing transverse momentum, 
with the first bin covering the range $\ptz \in [0,2]$~GeV. In this bin, the NNLO calculation of $\PZ$ production at finite transverse momentum 
diverges, and would need to be combined with the N$^3$LO three-loop contribution to inclusive $\PZ$ production, which is beyond the scope 
and aims of this study. 

\begin{figure}
  \centering
  \includegraphics[width=.82\linewidth]{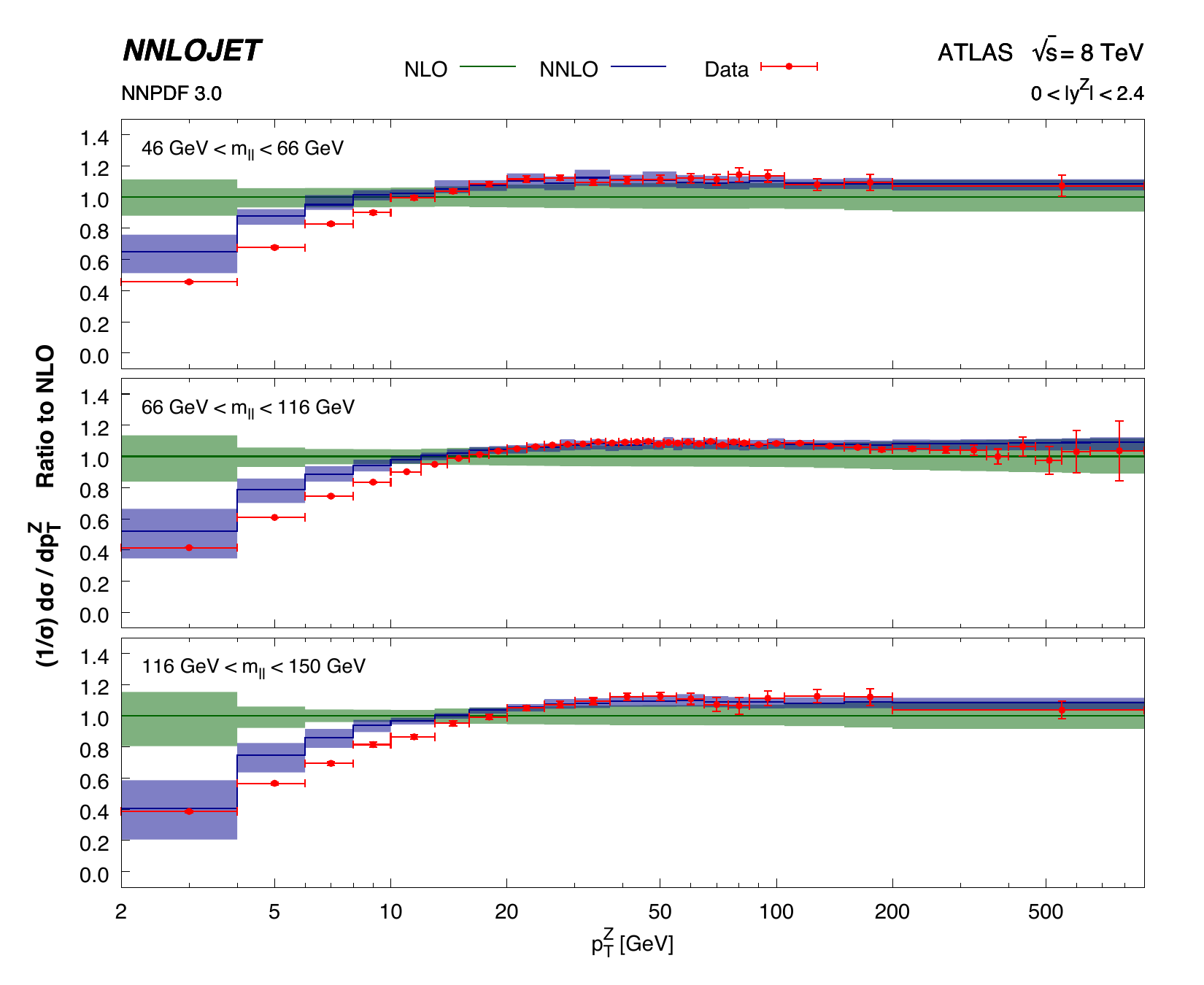}
  \caption{Normalized transverse momentum distribution differential in $\mll$ at NLO and NNLO compared to ATLAS data~\protect\cite{ptzATLAS}. The distribution is normalised to the NLO prediction.  The green bands denote the NLO prediction with scale uncertainty and the blue bands show the NNLO prediction
with scale uncertainty.}
  \label{fig:ptz:mll}
\end{figure}
\begin{figure}
  \centering
\includegraphics[width=.82\linewidth]{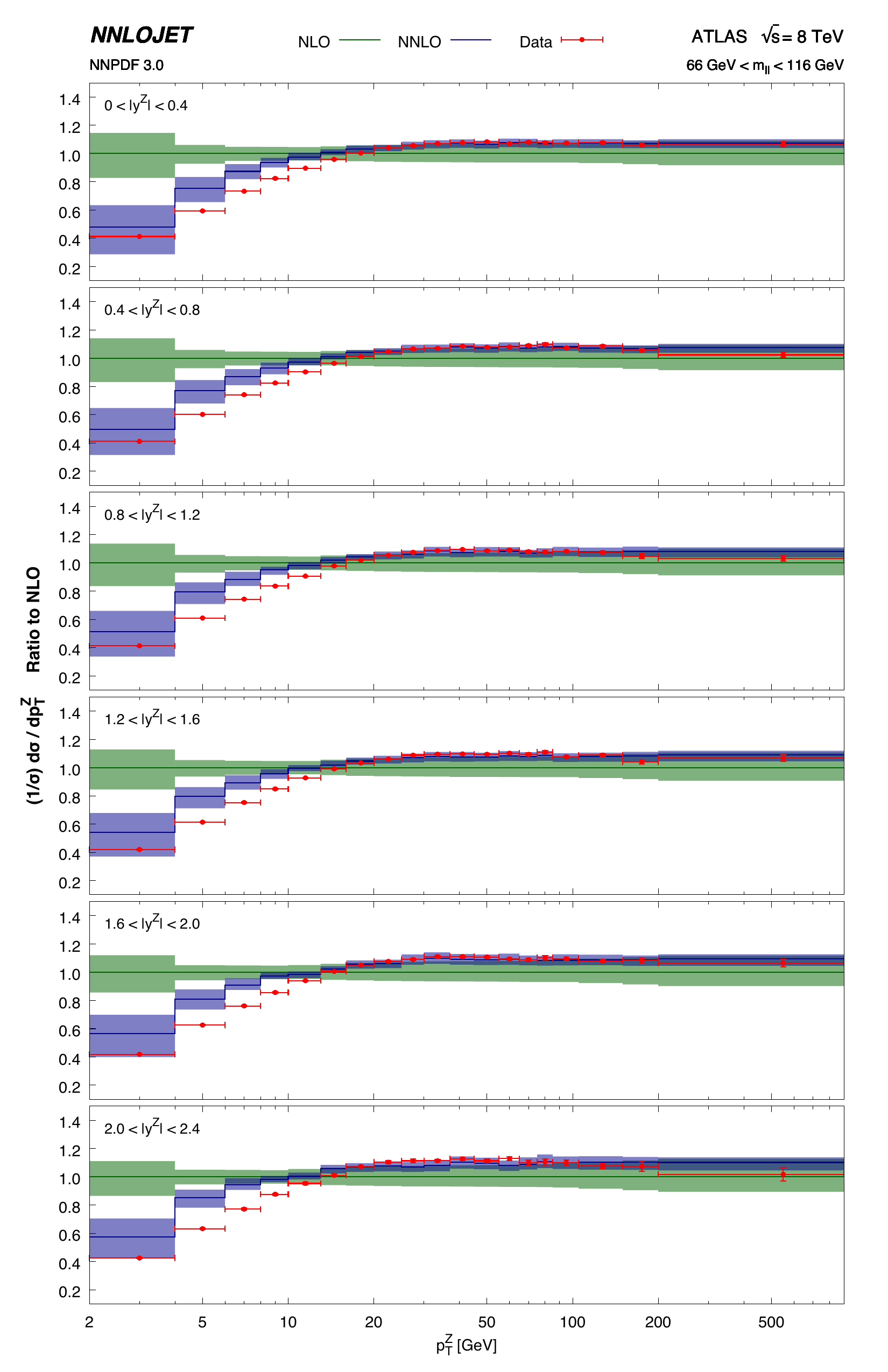}
  \caption{Normalized transverse momentum distribution differential in $y^{\PZ}$ for the 
  on-resonance bin at NLO and NNLO compared to ATLAS data~\protect\cite{ptzATLAS}. The distribution is normalised to the NLO prediction.  The green bands denote the NLO prediction with scale uncertainty and the blue bands show the NNLO prediction
with scale uncertainty.}
  \label{fig:ptz:yz}
\end{figure}

The measured $\ptz$ distribution is compared to the NLO and NNLO predictions in Figs.~\ref{fig:ptz:mll} and \ref{fig:ptz:yz}. 
%Note that the theory uncertainties shown in these figures are estimated using a 3-point scale variation instead of the 7-point variation described above by restricting to the case $\muf\equiv\mur$.
As already observed in Ref.~\cite{PTZus}, the NLO calculation does not describe the shape of the data below $\ptz\approx 40~\GeV$, while the 
NNLO calculation agrees with the data down to substantially lower values of $\ptz$. 
With the extended range in $\ptz$ that is covered in this study, we can now quantify this agreement in observing that the shape of 
the data is well-reproduced by the NNLO calculation down to $\ptz\approx 15~\GeV$. A deviation for lower values of $\ptz$ is expected due to 
the onset of large logarithmic corrections proportional to powers of $\ln(\ptz/\mll)$ in all orders in the strong coupling, which necessitates 
logarithmic resummation.  
The kinematical resolution below this value is however insufficient 
to resolve and quantify this  potential deviation in detail. In this region, the determination of $\ptz$ is limited by the 
experimental lepton
energy resolution. A more detailed picture can be gained from the distribution in the 
$\phistar$ variable, which is determined using the lepton angles rather than their  energies. This distribution is discussed in detail in 
the next two subsections.

%------------------------------------------------
\subsection{The large \texorpdfstring{$\phistar$}{phi*} region}
\label{sec:high-phi}
%------------------------------------------------

\begin{figure}
  \centering
  \includegraphics[width=.82\linewidth]{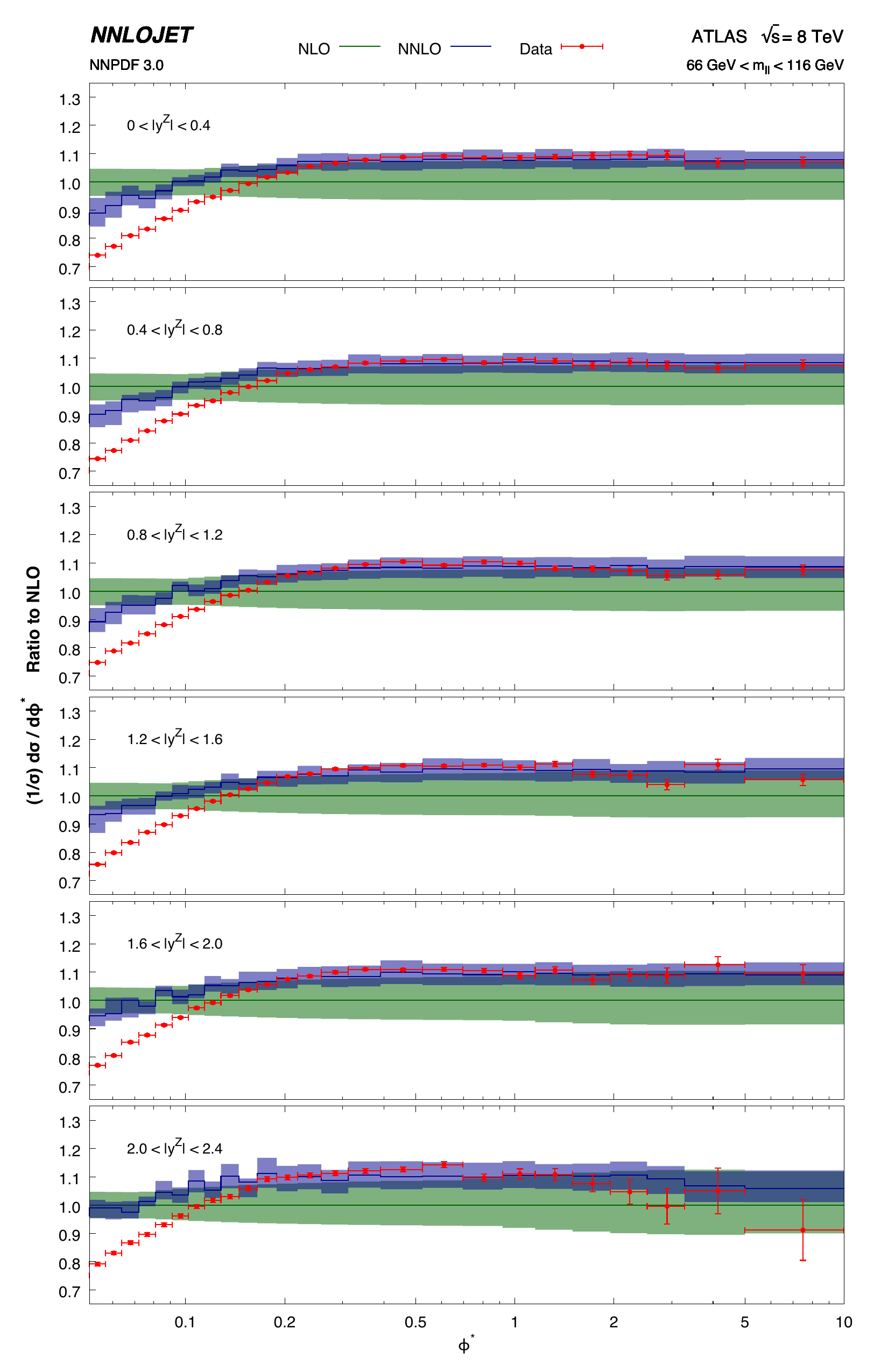}
  \caption{
    The $\phistar$ distribution for $\phistar > \phistarlarge$ for the on-resonance region $66~\GeV<\mll<116~\GeV$ in 6 different rapidity slices.
The distribution is normalised to the NLO prediction.  The green bands denote the NLO prediction with scale uncertainty and the blue bands show the NNLO prediction
with scale uncertainty.
   }
  \label{fig:phi_star_on_res}
\end{figure}

We first consider the region of large $\phistar$ values: $\phistar > \phistarlarge$, where one expects to see the transition 
between the fixed order behaviour in the bulk of 
the distribution and the onset of large logarithmic corrections at the lower end.

We first consider the region of large $\phistar$ values: $\phistar > \phistarlarge$
where one expects to see the transition 
between the fixed order behaviour valid to describe the bulk of the distribution 
and the onset of large logarithmic corrections needed at the lower end of this 
region.
Figure~\ref{fig:phi_star_on_res} shows the ratio of the normalised fixed-order predictions to the NLO prediction for the $\phistar$ distribution 
for each of the six rapidity slices in the on-resonance $\mll$-bin.
For $\phistar \gapprox 0.2$,  the NLO predictions systematically undershoot the data points by almost $5$--$10\%$.  This is reminiscent of the behaviour of the NLO prediction for the $\ptz$ distribution. The NNLO corrections are positive in this region and lead to a significant improvement in the theory--data comparison. Moreover, the residual scale uncertainty is greatly reduced by moving from NLO to NNLO. Below $\phistar \approx 0.2$ the 
shape of the NLO prediction quickly deviates away from the data points by more than $30$--$40\%$ for $\phistar \lesssim \phistarlarge$.
This divergent behaviour is tamed by the inclusion of the NNLO corrections where the shape of the
distribution shown by the data points is
better approximated  by the theory curve down towards smaller values of $\phistar$. It is however noted that even the 
NNLO terms, despite reproducing the tendency of the data, are insufficient to fully account for the quantitative 
behaviour of the data at lower $\phistar$. 

\begin{figure}
  \centering
  \includegraphics[width=.78\linewidth]{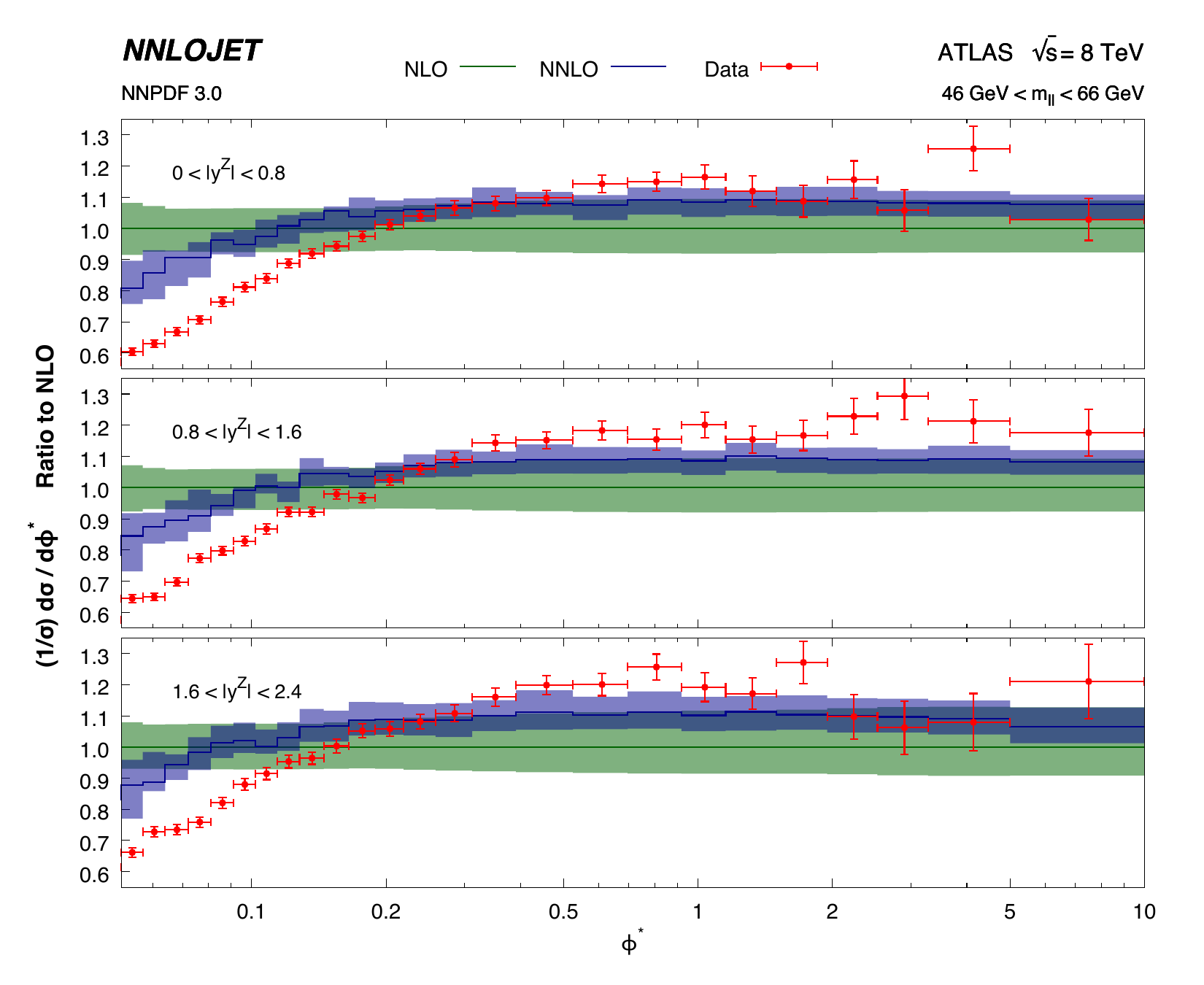} \\
  \includegraphics[width=.78\linewidth]{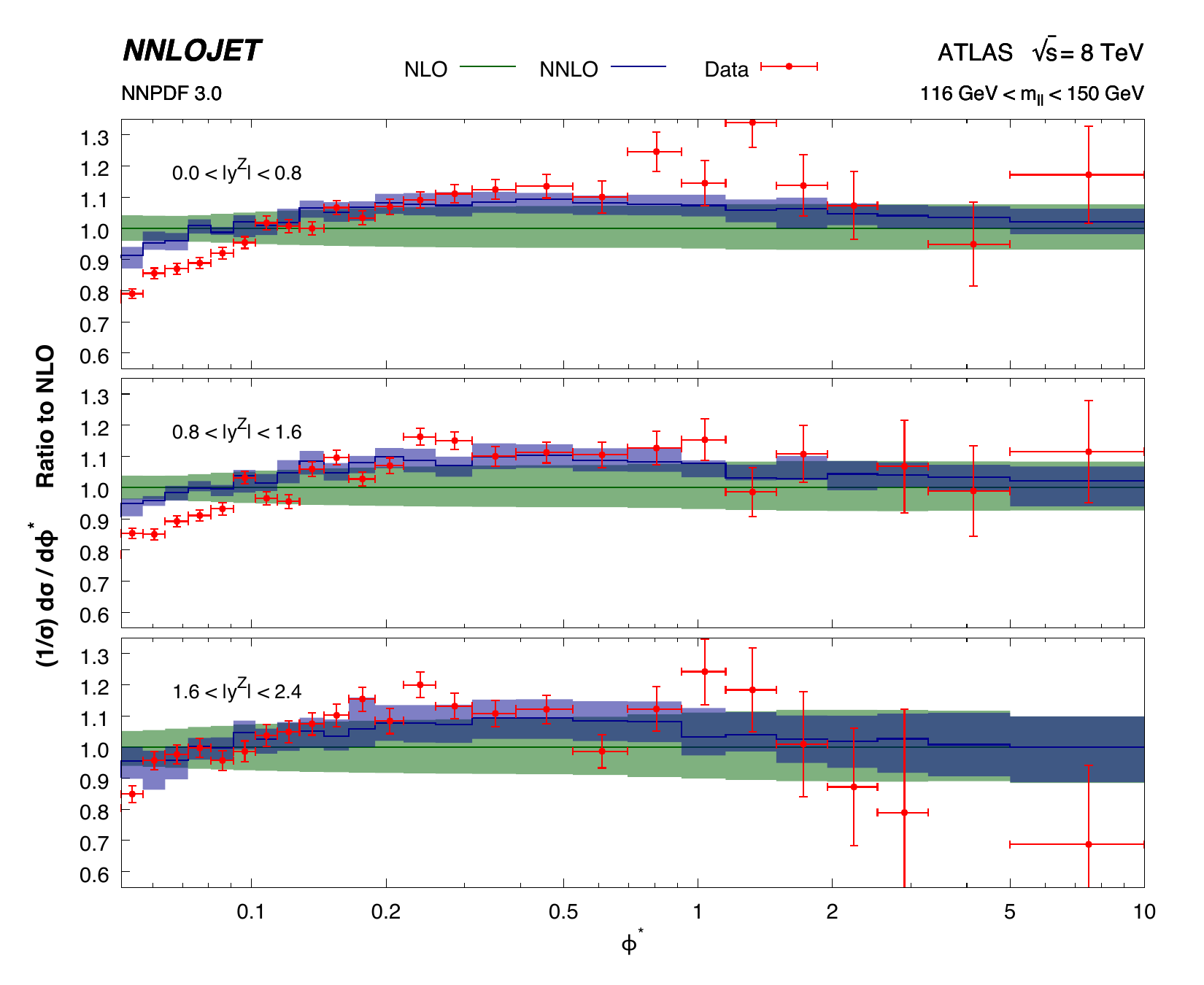}
  \caption{
    The $\phistar$ distribution for $\phistar > \phistarlarge$ for the two off-resonance $\mll$ bins,  $46~\GeV<\mll<66~\GeV$ (top) and $116~\GeV<\mll<150~\GeV$ (bottom) 
in 3 different rapidity slices.
The distribution is normalised to the NLO prediction.  The green bands denote the NLO prediction with scale uncertainty and the blue bands show the NNLO prediction with scale uncertainty.
  }
  \label{fig:phi_star_off_res}
\end{figure}

Figure~\ref{fig:phi_star_off_res} shows the \phistar\ distribution for the two off-resonance $\mll$ bins where the top and bottom plots in the figure correspond to the low-mass and high-mass bins, respectively. 
Due to lower event rates away from the $\PZ$-boson resonance region, the statistical errors on the experimental data points are significantly larger than the on-resonance results of Fig.~\ref{fig:phi_star_on_res}.
The qualitative picture, however, is very similar with a better theory--data agreement in the high-$\phistar$ region and improvements to the shape for  lower values of $\phistar$.

%------------------------------------------------
\subsection{The small \texorpdfstring{$\phistar$}{phi*} region}
\label{sec:low-phi}
%------------------------------------------------

At smaller values of $\phistar$, we enter the domain where the fixed-order perturbative prediction breaks down as large logarithms 
become important. 
 In this 
kinematical limit, $\phistar$ and $\ptz$ are closely related, as can be seen from the kinematical considerations in 
Section~\ref{sec:phistardef} above. These suggest the following approximate relation:
\begin{equation}
  \phistar \approx \ptz/\mll ,
 \label{eq:phiapprox}
\end{equation}
which can be used as guidance in the comparison of these distributions in the
region of low-$\ptz$ and low-$\phistar$.
The behaviour of the observable $\phistar$ in this region and the correspondence~\eqref{eq:phiapprox} have also been studied in refs.~\cite{BanfiphistarLHC,phistarlimit,atbanfi} in the context of logarithmic resummation. 

\begin{figure}
  \centering
  \includegraphics[width=.78\linewidth]{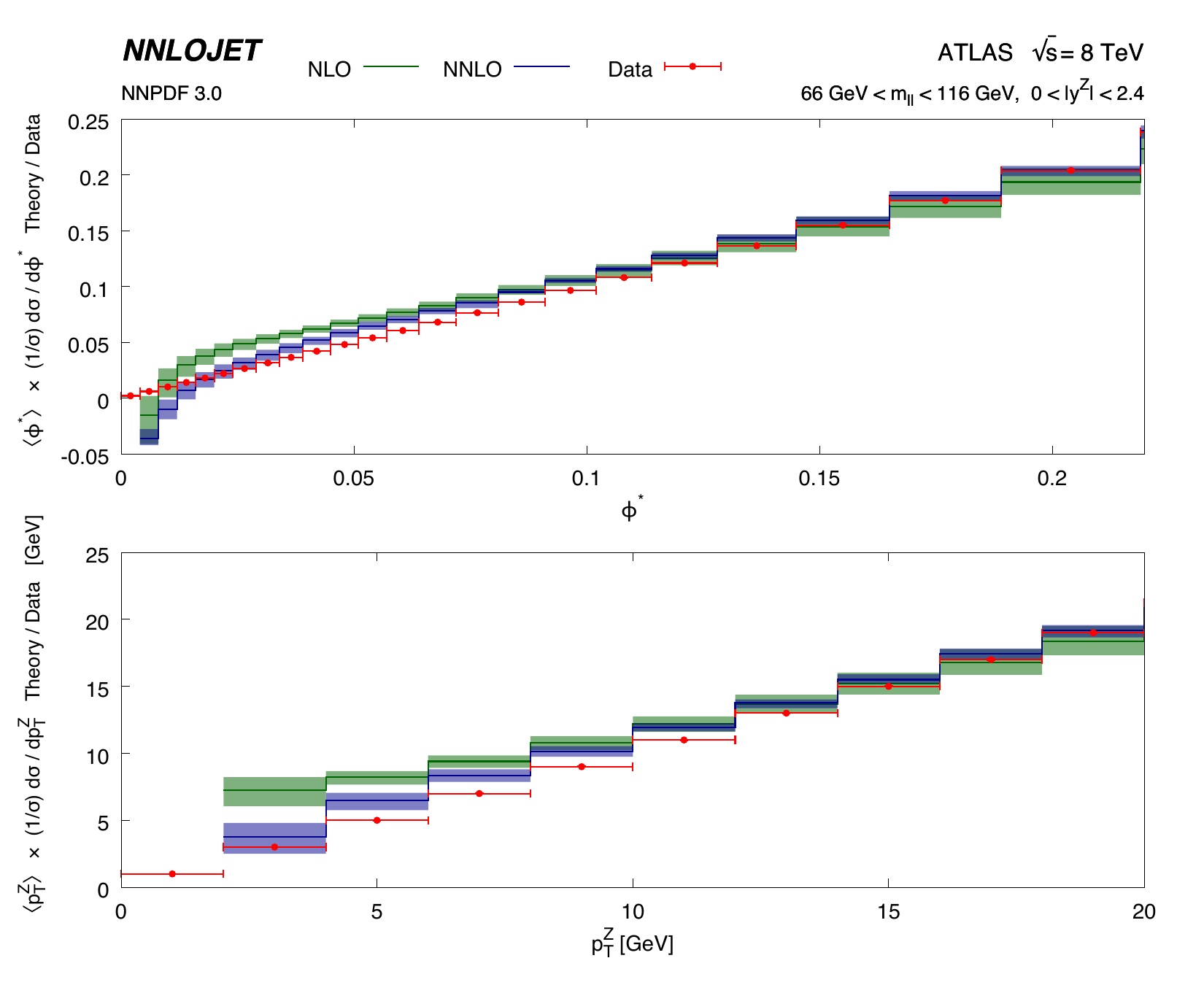}
  \caption{
 The $\phistar$ distribution for $\phistar > \phistarcut$ for the on-resonance mass bin $66~\GeV<\mll<116~\GeV$.
The distribution is normalised to the experimental data.  The green bands denote the NLO prediction with scale uncertainty and the blue bands show the NNLO prediction with scale uncertainty.    
  }
  \label{fig:lowphi_star_on_res}
\end{figure}
\begin{figure}
  \centering
  \includegraphics[width=.78\linewidth]{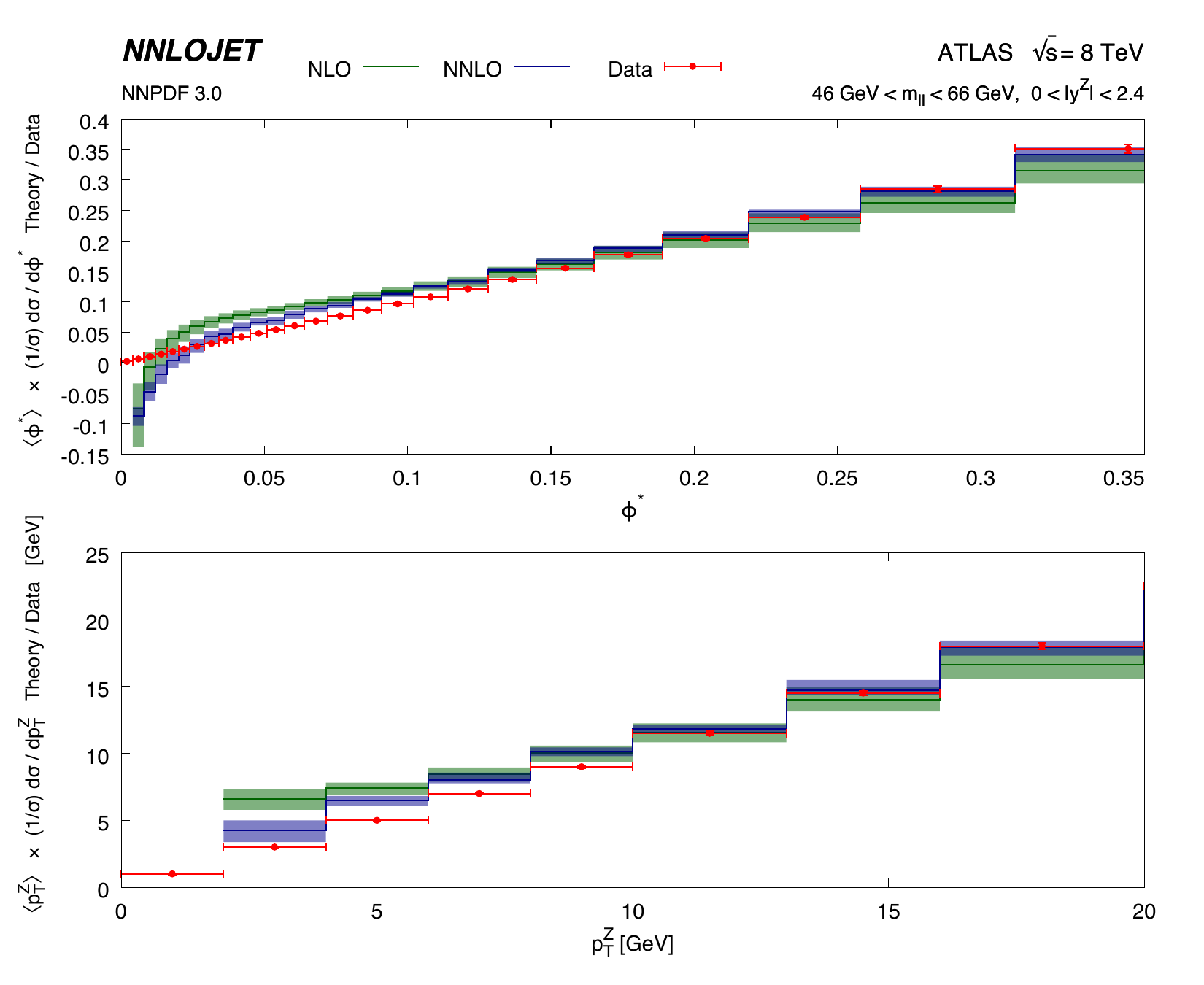}
  \caption{
 The $\phistar$ distribution for $\phistar > \phistarcut$ for the below resonance mass bin $46~\GeV<\mll<66~\GeV$.
The distribution is normalised to the experimental data.  The green bands denote the NLO prediction with scale uncertainty and the blue bands show the NNLO prediction
with scale uncertainty.    
  }
  \label{fig:lowphi_star_off_res_low}
\end{figure}
\begin{figure}
  \centering
  \includegraphics[width=.78\linewidth]{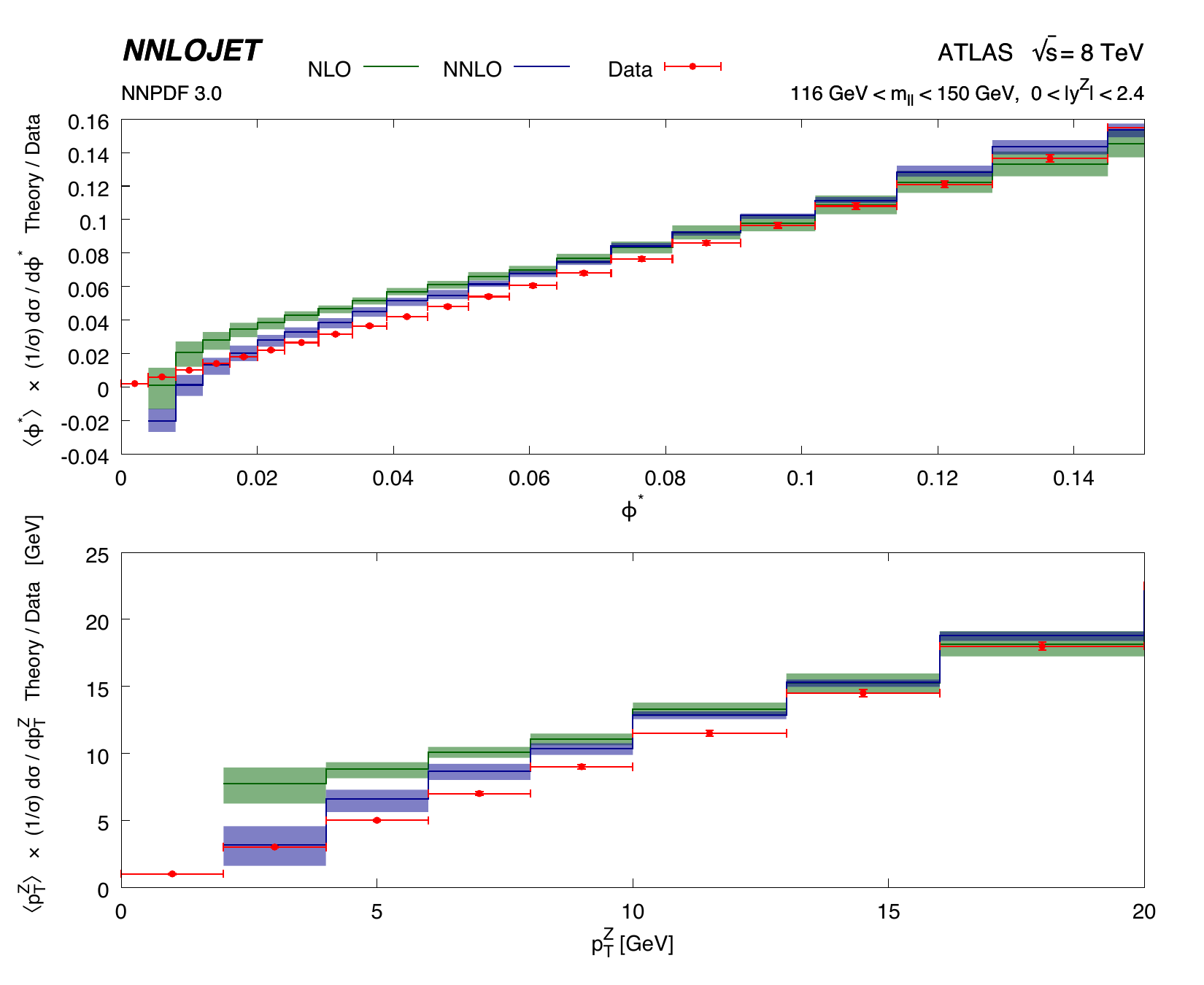}
  \caption{
 The $\phistar$ distribution for $\phistar > \phistarcut$ for the above resonance mass bin $116~\GeV<\mll<150~\GeV$.
The distribution is normalised to the experimental data.  The green bands denote the NLO prediction with scale uncertainty and the blue bands show the NNLO prediction
with scale uncertainty.    
  }
  \label{fig:lowphi_star_off_res_high}
\end{figure}

To test the range of validity of eq.~\eqref{eq:phiapprox} in the low-$\ptz$, low-$\phistar$ region, 
we superimpose the infrared regions of these distributions (for the three mass bins)
in Figs.~\ref{fig:lowphi_star_on_res}--\ref{fig:lowphi_star_off_res_high}.
For better visibility over the kinematical range, we show
\begin{equation*}
	\langle\mathcal{O}\rangle_\text{bin} \times \frac{\left. \frac{1}{\sigma}\, \frac{\rd\sigma}{\rd\mathcal{O}} \right\rvert_\text{Theory}}{\left. \frac{1}{\sigma}\, \frac{\rd\sigma}{\rd\mathcal{O}} \right\rvert_\text{Data}} 
	\quad\text{with}\quad 
	\mathcal{O} = \phistar,\;\ptz ,
\end{equation*}
i.e.\ the ratio of normalised distributions weighted by the central bin values.
The $\ptz$ range is fixed to $[0, 20]~\GeV$, while the 
$\phistar$ range is chosen according to Eq.~\eqref{eq:phiapprox} for each mass bin, using the central value of $\mll$. 
The first bins contain the zero value and are not accessible by a fixed-order calculation of the $\ptz$ or $\phistar$ distributions, which diverges there. 

First and foremost, we observe the substantially higher experimental resolution in $\phistar$: in the region covered by $(9+1)$ bins in $\ptz$, the $\phistar$ distribution contains 
$(22+1)$ bins on-resonance, $(25+1)$ bins below-resonance and $(19+1)$ bins above-resonance. This reflects the much better experimental resolution of the low $\ptz$ region afforded by the $\phistar$ variable.

In these figures, we observe that the NNLO predictions deviate from the data at larger 
values of $\ptz$ or $\phistar$ than at NLO, which is mainly due to the considerable reduction in the scale
uncertainty: the larger NLO uncertainty band allows to remain compatible with the data. In terms of describing 
the shape of the data, the NNLO predictions better reproduce the general tendency of the data in this region, 
however without providing a detailed quantitative agreement. 
For both distributions, we also recall that the NNLO corrections improve the description of the shape 
of the distribution at larger values of $\ptz$ or $\phistar$ (outside the plotting range here). 
By comparing the departure points between theory and data for both observables, we are able to test the approximate kineamtic relation $\ptz \sim \mll\,\phistar$.
The results are summarised in Table~\ref{tab:depart} and show that the relation is indeed fulfilled to a reasonable level.
\begin{table}
  \centering
  \small
  \begin{tabular}{ @{\enskip}c@{\quad} r@{--}l c c c }
  \toprule
  & \multicolumn{2}{@{\quad}c@{\quad}}{$\mll$-bin $\;[\GeV]$} &
  $\phistardep$ &
  $\ptzdep$ &
  $\langle \mll \rangle\, \phistardep$ \\
  \cmidrule(r{2pt}){2-3} \cmidrule(l{2pt}){4-6}
  \multirow{3}{*}{\rotatebox[origin=c]{90}{NLO}}
  & \qquad 46  & 66  & $\sim 0.14$ & $\sim  9~\GeV$ & $ 8~\GeV$ \\
  &        66  & 116 & $\sim 0.11$ & $\sim 11~\GeV$ & $10~\GeV$ \\
  &        116 & 150 & $\sim 0.08$ & $\sim 12~\GeV$ & $11~\GeV$ \\
  \cmidrule(r{2pt}){2-3} \cmidrule(l{2pt}){4-6}
  \multirow{3}{*}{\rotatebox[origin=c]{90}{NNLO}}
  & 46  & 66  & $\sim 0.17$ & $\sim 10~\GeV$ & $10~\GeV$ \\
  & 66  & 116 & $\sim 0.14$ & $\sim 15~\GeV$ & $13~\GeV$ \\
  & 116 & 150 & $\sim 0.13$ & $\sim 20~\GeV$ & $17~\GeV$ \\
  \bottomrule
  \end{tabular}
  \caption{Values of $\phistardep$ and $\ptzdep$ for the three mass windows corresponding to the values of $\phistar$ and $\ptz$ where the fixed order predictions of the distributions start to deviate from the experimental data.}
  \label{tab:depart}
\end{table}

%------------------------------------------------
\section{Summary and conclusions}
\label{sec:summary}
%------------------------------------------------

We have studied the NNLO QCD corrections to the Drell--Yan production of lepton pairs at small transverse momentum, inclusive over the hadronic final state.  There are two relevant observables, $\ptz$ and $\phistar$.  From the experimental point of view, $\phistar$ relies on knowing the lepton direction and can be measured more precisely at low transverse momentum than $\ptz$ where the momenta of the final state leptons largely cancel. 

Our calculation is performed using the parton-level Monte Carlo generator \NNLOJET which
implements the antenna subtraction method for NNLO calculations of hadron collider observables. It extends our earlier calculations of $\PZ/\gamma^*+\jet$ production~\cite{ZJNNLOus} and $\PZ/\gamma^*$ production at large transverse momentum~\cite{PTZus}. We have performed a thorough comparison
of theory predictions to the 8 TeV Run 1 data of the LHC for cross sections defined over a
fiducial region of lepton kinematics from the ATLAS~\cite{ptzATLAS} collaboration.

At large values of $\phistar$, we
observe that the NNLO corrections to the distribution normalised to the inclusive NNLO dilepton cross section are moderate and positive, resulting in an excellent agreement between data and theory. This agreement holds across the three $\mll$ bins and for all slices in $\yz$. 

In the small transverse momentum region, we expect the fixed order calculation to break down due to the emergence of large logarithmic corrections of the form $\alphas^n\log^{2n-1}(\ptz)$ or  $\alphas^n\log^{2n-1}(\phistar)$.  
The inclusion of NNLO contributions partly captures this behaviour, and provides qualitative improvements in the 
description of the shape of the data. A detailed quantitative comparison, also taking into account the 
considerably lower scale uncertainty of the predictions at NNLO, shows that deviations between data and 
theory start to become visible at comparable values of $\ptz$ or $\phistar$ both at NLO and NNLO. 
We also showed that these breakdown points satisfy an approximate relationship 
$\ptz \sim \mll \phistar$ that is motivated by approximating the process kinematics in this region. 

The NNLO corrections improve the perturbative description of the $\ptz$ and $\phistar$ distributions by reducing the theoretical scale uncertainty and better accounting for the shape of the data especially at large values of these 
variables. We anticipate that this calculation
will allow
a consistent inclusion of the precision data on the Z transverse momentum distribution
into NNLO determinations of parton distributions and the strong coupling constant.

\subsection*{Note added}
The numerical results of the published JHEP version of this paper \cite{Gehrmann-DeRidder:2016jns}
contained an implementation error that was uncovered only at a later stage, when performing in-depth
validations of the numerical code against the fixed order NNLO expansion of the newly derived~\cite{Bizon:2018foh} 
resummation of third-order logarithmic corrections (N${}^\text{3}$LL). An erratum correcting this error has been 
published in JHEP.  The present arXiv version (v3) of the manuscript integrates the content of the erratum. 

We thank Pier Francesco Monni and Alexander Karlberg for many discussions and 
in-depth comparisons which have led us to the 
identification of the implementation error.

\acknowledgments

The authors thank
Xuan Chen, Juan Cruz-Martinez, James Currie, Jan Niehues and Joao Pires for useful discussions and
their many contributions to the \NNLOJET\ code.
We gratefully acknowledge the computing resources provided by the WLCG through the GridPP Collaboration.
This research was supported in part by the National Science Foundation under Grant NSF PHY11-25915,
in part by the Swiss National Science Foundation (SNF) under contracts 200020-162487 
and CRSII2-160814, in part by
the UK Science and Technology Facilities Council, in part by the Research Executive Agency (REA) of the European Union under the Grant Agreement PITN-GA-2012-316704  (``HiggsTools'') and the ERC Advanced Grant MC@NNLO (340983).

\end{document}